\documentclass[prd,aps,superscriptaddress,floatfix,nofootinbib,eqsecnum]{revtex4-2}

\pdfoutput=1

\usepackage{amsfonts}
\usepackage{amsmath}
\usepackage{amssymb}
\usepackage{bm}
\usepackage{dcolumn}
\usepackage{graphicx}
\usepackage[latin1]{inputenc}
\usepackage{latexsym}
\usepackage{rotating}
\usepackage{hyperref}
\usepackage{subfigure}
\usepackage{color}
\usepackage{changes}
\usepackage{verbatim}
\usepackage{comment}
\usepackage{empheq}
\usepackage{csquotes}
\usepackage{physics}
\usepackage{float}
\usepackage{amsmath,latexsym}
\usepackage{booktabs}  

\begin{document}

\title{Generalized uncertainty principle distorted quintessence dynamics
}

\author{Gaurav Bhandari}\email{bhandarigaurav1408@gmail.com}\affiliation{Department of Physics, Lovely Professional University, Phagwara, Punjab, 144411, India}

\author{S. D. Pathak}\email{shankar.23439@lpu.co.in}\affiliation{Department of Physics, Lovely Professional University, Phagwara, Punjab, 144411, India}

\author{Manabendra Sharma}\email{sharma.man@mahidol.ac.th}\affiliation{Centre for Theoretical Physics and Natural Philosophy, Nakhonsawan Studiorum for Advanced Studies, Mahidol University, Nakhonsawan, 60130, Thailand}

\author{Anzhong Wang}\email{Anzhong$_$Wang@baylor.edu}\affiliation{GCAP-CASPER, Physics Department, Baylor University, Waco, TX 76798-7316, USA}

\begin{abstract} In this paper, we invoke a generalized uncertainty principle (GUP) in the symmetry-reduced cosmological  Hamiltonian for a universe driven by a quintessence scalar field with potential. Our study focuses on semi-classical regime. In particular, we derive the GUP-distorted Friedmann, Raychaudhuri, and the Klein-Gordon equation. 
This is followed by a systematic analysis of the qualitative dynamics for the choice of potential $V(\phi)= V_0 \sinh^{-n}{(\mu \phi)}$. This involves constructing an autonomous dynamical system of equations by choosing appropriate dynamical variables, followed by a qualitative study using linear stability theory. Our analysis shows that incorporating GUP significantly changes the existing fixed points compared to the limiting case without quantum effects by switching off the GUP. 
\end{abstract}

\maketitle


\tableofcontents

\section{Introduction} \label{Introduction}

The discovery of Einstein's general theory of relativity gave rise to many fields of research.
Since gravity becomes predominant at large distances, general relativity (GR) offers  a robust  mathematical framework for constructing different cosmological models. Over the last hundred years GR has provided explanations to many phenomena of the cosmos including an explanation of perihelion precession of mercury \cite{rana1987investigation}, deflection of light when passing through massive bodies \cite{genov2009mimicking}, gravitational redshift of light \cite{wojtak2011gravitational} and black holes \cite{o1,ku1}.

The modern cosmology stands on the basic principles of homogeneity and isotropy dubbed the cosmological principle. The uniform and isotropic expansion of the universe is described by the Friedmann-Lemaitre-Robertson-Walker (FLRW) metric, which, when applied to General Relativity (GR), gives rise to the standard model (SM) of cosmology. The prediction of cosmological microwave background (CMB) radiation is one of the remarkable successes of the standard model of cosmology.

While SM is successful in explaining many aspects of the universe, it is confronted with some serious drawbacks like horizon, flatness, and entropy problems\cite{albrecht1982cosmology,guth1981inflationary,liddle1992cobe,kinney2009tasi,riotto2002inflation}.
In order to address these issues A. Guth in 1981 proposed
a nearly exponential expansion of the universe at an early phase called inflation, in his seminal work \cite{A1,A2}. However, inflation does not resolve the problem of initial singularity in such classical cases as discussed in \cite{pan1,bro1,zhu1,sh5,bat1,ca1}. Nevertheless, inflationary theory not only resolves the problems of SM but also predicts the large-scale structure of the universe. Despite the apparent homogeneous and isotropic at large scale \cite{A.A.Starobinsky}, minute fluctuation of the order of $10^{-5}$ is observed in CMB radiation.  These small-scale fluctuations enable us to use perturbative theory, where the zeroth order background is still homogeneous and isotropic FLRW metric. The inhomogeneity is governed by leading order correction terms. One of the main causes of spacetime perturbation is the quantum fluctuation of matter content. These fluctuations within the matter field impact the metric, causing energy and matter density to clump together, ultimately leading to the formation of the large structures we observe today. In perturbative theory, the background metric is treated classically while quantizing the first-order correction term in linearized gravity theory. However, in a genuine quantum theory of gravity, one must account for the quantum nature of the background metric at scales approaching the Planck region. \cite{guth1982fluctuations,linde1982new,bardeen1983spontaneous,ashtekar2005quantum,sing1,ashtekar2009loop}.

The persistent lack of a complete understanding of early universe cosmology stems from the absence of consistent candidates for quantum gravity. Moreover, the challenge of reconciling two mathematically incompatible frameworks, general relativity (GR) and quantum mechanics (QM), continues to elude our current comprehension, as discussed in \cite{c1,s1,s2,h1}.

In recent times, various candidates for quantum gravity have emerged, each possessing its own advantages and disadvantages. Among these is loop quantum gravity (LQG)\cite{rovelli1998strings,sharma2019background,garay1995quantum,as1,ha1,sh1}, string theory(ST) \cite{veneziano1986stringy,witten1996reflections,scardigli1999generalized,gross1988string,amati1989can,yoneya1989interpretation,s3}, doubly special relativity (DSR)  \cite{double1, gh1}. In all of these theories, the existence of a minimum length scale on the order of the Planck length is evident. This minimal length scale constitutes the length of the string itself in string theory, and the existence of a minimal area gap in Loop Quantum Gravity (LQG).\cite{ashtekar2021short,rovelli2008loop,ra1,zhu2017pre,li2018qualitative}.

In the ongoing absence of a consistent theory of quantum gravity, the existence of a minimal length serves as motivation for various approaches to quantizing gravity. One such approach is the Generalized Uncertainty Principle (GUP), which extends the Heisenberg Uncertainty Principle (HUP) by introducing a momentum correction term, as proposed in Loop Quantum Gravity (LQG)\cite{m1,si1}.
In the non-relativistic case, the HUP serves as the departure point from classical mechanics to standard quantum mechanics. It asserts the incompatibility between position and momentum operators, illustrating the inherent imprecision in measuring one when the other is precisely known. Similarly, in the relativistic case, the GUP is anticipated to serve a similar role.

The GUP, indeed offers an alternative approach aimed at incorporating quantum corrections into cosmological dynamics, particularly for studying the early universe. In this framework, classical points in spacetime are regarded as probability densities associated with basis vectors. Additional geometric fluctuations give rise to the Extended Generalized Uncertainty Principle (EGUP)\cite{sc1,kempf1994quantum,kempf1997quantum}. 

Considering quantum fluctuations in space-time leads to the theory of GUP, which imposes limitations on the measurement of position and momentum. These uncertainties in the geometry of space contribute to the increased uncertainty in measuring the position and momentum of a particle \cite{tawfik2015review,lake2021generalised,PhysRevD.85.104029,ADLER_1999}. The proposition that gravity could influence the uncertainty principle was initially put forth by Mead in 1964 \cite{mead1964possible}.

The GUP has demonstrated remarkable efficacy in elucidating the dynamics of the universe. For example, it provides valuable insights into the genesis of magnetic fields with strengths at the scale of microgauss, as observed in intergalactic regions when accounting for the GUP \cite{so1,so2,so3}. Furthermore, the GUP exhibits promising applications in the domain of black hole mechanics, particularly in the context of radiating small black holes in accordance with the Bekenstein entropy relation and Hawking radiation, as elaborated in \cite{bo1,bt1,bo2,k1}. Considering this perspective, it is important to examine different cosmological models through the lens of GUP to determine whether the effects of GUP enhance our comprehension and point towards intriguing avenues concerning the dynamic history of the universe.

The quintessential inflation model comprises solely one classical scalar field, referred to as the inflaton, within the standard inflation paradigm. This model offers a unified framework for elucidating the acceleration of both early and late times through the application of a singular scalar field. Frequently observed within quintessence inflation models are attractor behavior, wherein the dynamics of the scalar field yield specific solutions irrespective of the initial conditions, thereby mitigating the issue of fine-tuning. The incorporation of quintessence into inflationary theory marked a pioneering endeavor by Peebles and Vilenkin, as detailed in their seminal paper \cite{pe1}.  Furthermore, this model also offers specific predictions regarding primordial density perturbations, which can be compared with observational data from the Cosmic Microwave Background (CMB). Through the study of the quintessence scalar field, we can delve into the physics of early times and explore the mechanisms behind cosmic inflation \cite{co1,rh1,lu1,dev,mmv}.

In this paper, our focus is on the minimally coupled quintessence scalar field (canonical scalar field) to examine the overall dynamics in the presence of GUP. The paper is organized as follows:
In Sec.\ref{GUP}  we construct the formulation of the GUP-corrected Hamiltonian by considering the Einstein-Hilbert action with a minimally coupled quintessence scalar field featuring an arbitrary potential. We obtain the GUP-corrected Friedmann, Raychaudhuri and the Klein-Gordon equation in Sec.\ref{GUP deformed}. Later, in Sec.\ref{dsa},  we employ techniques of dynamical system analysis to extract qualitative information about the newly introduced GUP-distorted system and perform linear stability analysis. We confine our examination to a specific choice of potential of the form $V(\phi)= V_0 \sinh^{-n}{\mu \phi}$.

\section{GUP-modified background cosmology}
The Friedmann--Lemaitre--Robertson--Walker (FLRW) metric for the spatially flat universe ($k=0$) which guarantees homogeneity and isotropy is given by
\begin{equation}\label{metric}
ds^2= -N^2(t)dt^2 + a^2(t)\left[dr^2 +  r^2 d\Omega^2\right],
\end{equation}
The Einstein-Hilbert (EH) action incorporating the metric and matter field for the scalar field is given:
\begin{equation}\label{Action}
S_{EH}= \frac{1}{2\kappa}\int d^4x \sqrt{-g} R + \mathcal{L}_m,
\end{equation}
which gives $G_{\mu \nu}=T_{\mu \nu}$, where $G_{\mu \nu}$ is the Einstein tensor  and $T_{\mu \nu}$ is the energy-momentum tensor. Considering natural units we set $\kappa\equiv \frac{1}{8\pi G}$ equal to one for rest of the paper.

According to GR matter defines geometry (Mach principle) and this geometry dictates the flow of matter in the universe. This implies that the matter content of the universe determines the expansion of universe for a given epoch. The second approach is to modify the geometric sector of EH action Eq.(\ref{Action}) which gives rise to a plethora of models \cite{rat1,cal1,fen1}.

In this paper, our focus lies on the former approach, wherein we employ the quintessence scalar field to examine the dynamics of the universe's evolution. The main objective is to delve into the dynamics of the universe, where quantum gravity effects remain significant but not predominant.

We adopt the approach of modeling the quantum-corrected background evolution of the universe by incorporating the GUP directly into the cosmological Hamiltonian. In this section, we will review the method of constructing the equation of motion for the quintessence scalar field without GUP deformation.

The Lagrangian for quintessence scalar filled with arbitrary potential is
\begin{equation} \label{Lag}
\mathcal{L}= \frac{1}{2\kappa}\sqrt{|g|} \left[R+\kappa\{ (\partial_{\mu} \phi)(\partial_{\nu} \phi) -2V(\phi)\}\right] ,   
\end{equation}
for FLRW  universe with chosen signature as $(-,+,+,+)$, Lagrangian (\ref{Lag}) reduces to\cite{novosyadlyj2013quintessence},
\begin{equation}
\mathcal{L}= -3a\Dot{a}^2+a^3\left(\frac{\Dot{\phi}^2}{2}-V(\phi)\right).
\label{phantomlang}
\end{equation}
One can obtain Hamiltonian through the Legendre transformation, defined as $\mathcal{H} \equiv \dot{a}P_a+ \dot{\phi}P_{\phi}-\mathcal{L},$ to give classical Hamiltonian as
\begin{equation}\label{ClassicalHamiltonian}
\mathcal{H}=-\frac{P_a^2}{12a}+\frac{P_\phi^2}{2a^3}+a^3V(\phi).
\end{equation}

The classical Hamiltonian given by Eq.(\ref{ClassicalHamiltonian}) governs the background dynamics of a universe dominated by a quintessence scalar field with the following symplectic structure:
\begin{eqnarray}\label{Symplectic}
\{a,P_a\}   = 1,\quad
\{\phi, P_{\phi}\} = 1,
\end{eqnarray}
The conjugate momenta to $a$ and $\phi$ are denoted as $P_a\equiv \frac{\delta \mathcal{L}}{\delta \dot{a}}$ and $P_{\phi}=\frac{\mathcal{L}}{\delta \dot{\phi}}$ respectively. Therefore, the complete phase space comprises $(a,P_a,\phi, P_{\phi})$. While the symplectic algebra represented by Eq.(\ref{Symplectic}) captures the kinematical structure of the theory, the dynamical evolution is determined by the Poisson flow of the phase space variables with respect to the Hamiltonian $\mathcal{H}$.

\begin{eqnarray}\nonumber
\dot{a} &=& \{a, \mathcal{H} \}, \quad 
\dot{P_a} = \{P_a, \mathcal{H} \},\nonumber \\
\dot{\phi} &=& \{\phi, \mathcal{H} \}, \quad 
\dot{P_{\phi}} = \{P_{\phi}, \mathcal{H}\}.\label{poisson}
\end{eqnarray}
Using the above Eq.(\ref{poisson}) the Friedmann, Raychaudhuri and Klien-Gordon equations for the Quintessence scalar field are obtained as follows:
\begin{eqnarray}
&&3H^2=\frac{\dot \phi^2}{2}+V(\phi),\\
&& \Ddot{\phi}+3\Dot{\phi}\frac{\Dot{a}}{a}+\frac{dV(\phi)}{d\phi}=0,\\
\label{Vklein}
&& 2\frac{\Ddot{a}}{a^2}+ \left(\frac{\Dot{a}}{a}\right)^2= - \frac{\Dot{\phi}^2}{2} + V(\phi).
\end{eqnarray}

So far, our analysis has been purely classical. However, formulating the theory in canonical language paves the way for canonical quantization. In the following section, we will explore how GUP correction deforms the symplectic structure in question. Once this is understood, it is straight forward to obtain the quantum-corrected dynamics. 

\subsection{
Quintessence  scalar field in GUP-modified setup
}
\label{GUP}

\subsubsection{Classical formulation} \label{classicaltreatment}
The Einstein--Hilbert action with a minimally coupled scalar field and arbitrary potential is
\begin{equation}
S_{EH} = \int \sqrt{-g} \left[ \frac{1}{2\kappa}\left(R-V(\phi) \right) + \frac{1}{2}\partial_{\mu}\phi\partial_{\nu}\phi\right]d^4x,
\end{equation}

In the context of a maximally symmetric spacetime $ds^2= -N^2(t)dt^2+a^2(t)[dr^2+r^2d\Omega^2]$, N(t) represents the lapse function, the action corresponding to the FLRW background can be expressed as: 

\begin{equation}
S= V_0 \int dt \left[ -\frac{3a\dot{a}^2}{N}+a^3\left(\frac{\dot{\phi}^2}{2N}- N V \right)\right]. \label{cosmo}
\end{equation}

For our calculation within a non-compact FLRW spacetime, we introduce a fictitious volume $V_0$
to aid in our calculations. However, since it does not influence the dynamics, we can arbitrarily set it equal to 1 without any loss of generality. Additionally, for the rest of the paper, we adopt natural units where $\kappa=1$

Thus, the symmetry reduces Lagrangian from Eq.(\ref{cosmo}) is given by:
\begin{equation}\label{FreePhantomLagrangian}
\mathcal{L}=-\frac{3a\dot a^2}{N}+a^3\left(\frac{\dot\phi^2}{2N} - NV\right).
\end{equation}
\\
It is obvious that the Lagrangian, Eq.(\ref{FreePhantomLagrangian}), is devoid of $\dot{N}(t).$ and hence there is no dynamics in the lapse function $N(t)$,  $P_{N} \equiv \frac{\partial \mathcal{L}}{\partial \dot{N}}=0$ being a constant of motion. Therefore, the dynamics of the system are completely contained in the equation of motion for $(a, P_a, \phi, P_{\phi})$ governed by the Hamiltonian:
\begin{equation}
\mathcal{H}=N\left[-\frac{P_a^2}{12a}+\frac{P_\phi^2}{2a^3}+a^3V\right],\label{Hamiltonian}
\end{equation}
which is obtained from Eq.(\ref{FreePhantomLagrangian}) through the Legendre transformation.\newline
The Equation of motion can be obtained  using Eq.(\ref{poisson})
and by substituting the form of $P_a$ and $P_\phi$ so obtained from the Lagrangian  Eq.(\ref{FreePhantomLagrangian}).
The Raychaudhuri equation is
\begin{equation}
2\frac{\Ddot{a}}{a}+\left(\frac{\dot a}{a}\right)^2= -\frac{\dot \phi^2}{2}+V(\phi).
\end{equation}
The Friedmann equation is
\begin{equation}
3H^2=\frac{\dot \phi^2}{2}+V(\phi).
\end{equation}
Furthermore, the Klein-Gordon (KG) equation is
\begin{equation}
\Ddot{\phi}+3H\dot \phi +V,_\phi(\phi)=0.
\end{equation}

\subsubsection{GUP-distorted formulation} \label{GUP deformed}
In this subsection, we invoke the higher-order correction of uncertainty principle to the cosmological Hamiltonian. In order to achieve this, we start with choosing the phase space $x$ and $y$ which incorporate scale factor $a$ and scalar field $\phi$ as: 
\begin{equation}\label{CanonicalTransformation}
x= \frac{a^{3/2}}{\mu}\sinh(\mu \phi),  \hspace{0.5cm} y= \frac{a^{3/2}}{\mu}\cosh(\mu \phi),
\end{equation}
while dynamics remain the same. We note that the differentiation of Eq.(\ref{CanonicalTransformation}) gives
\begin{eqnarray}
\dot x =\frac{3}{2}\frac{a^{1/2}\dot a}{\mu} \sinh (\mu\phi)+ (a^{3/2}\dot\phi) \cosh(\mu\phi),\label{xdot}\\
\dot y = \frac{3}{2}\frac{a^{1/2}\dot a}{\mu} \cosh (\mu\phi)+ (a^{3/2}\dot\phi) \sinh(\mu\phi).\label{ydot}
\end{eqnarray}
Now, multiplying both sides of Eq.(\ref{xdot}) by $\sinh(\mu\phi)$ and Eq.(\ref{ydot}) by $\cosh(\mu\phi)$ and subtracting  them together gives
\begin{equation}
\dot y \cosh(\mu\phi)- \dot x \sinh(\mu\phi)= \frac{3}{2}\frac{a^{1/2}}{\mu}\dot a.\label{Eq1}
\end{equation}
Next by squaring both sides of Eq.(\ref{Eq1}), followed by dividing the expression by $2N$ and considering $\mu=\sqrt{3/8}$ one obtains,
\begin{equation}
\dot y^2 \cosh^2(\mu\phi)+\frac{\dot x ^2 \sinh^2(\mu\phi)-2\dot x \dot y \sin(\mu\phi)\cos(\mu\phi)}{2N}=\frac{3a\dot a^2}{N}. \label{FEq}
\end{equation}
Instead, multiply both sides of  Eq.(\ref{xdot}) by $\cosh(\mu\phi)$ and Eq.(\ref{ydot}) by $\sinh(\mu\phi)$ and subtracting
\begin{equation}
\dot x \cosh(\mu\phi)-\dot y \sinh(\mu\phi)= a^{3/2} \dot \phi.\label{Eq2}
\end{equation}
This is followed by squaring Eq.(\ref{Eq2}) and dividing it by $2N$ to obtain
\begin{equation}
\frac{\dot x ^2 \cosh^2(\mu\phi)+ \dot y^2 \sinh^2(\mu\phi)-2\dot x \dot y \cos(\mu\phi)\sin(\mu\phi)}{2N}= \frac{a^3\dot \phi^2}{2N}. \label{terms}
\end{equation}
In subtracting, by considering the squares of the components in  Eq.(\ref{CanonicalTransformation}), we have:
\begin{align*}
y^2-x^2 = \frac{a^3}{\mu^2} \sin^2(\mu\phi)+ \frac{a^3}{\mu^2} \cos^2(\mu\phi)
\end{align*}
\begin{equation}
\implies y^2-x^2 = \frac{a^3}{\mu^2}= \frac{8a^3}{3}.\label{Circle}
\end{equation}

After converting all the terms in cartesian pair (x,y), we now return to the question of dynamics. Using Eq.(\ref{FEq}) Eq.(\ref{terms}) and Eq.(\ref{Circle})  we can determine the 
 form of the Lagrangian in configuration coordinates $(x,y,\dot{x},\dot{y})$ to be
\begin{equation}
\mathcal{L}= \left[\frac{\dot x^2 + \dot y^2}{2N} - \frac{3}{8}(y^2-x^2)V(\phi) N \right].
\end{equation}
Using the above langrangian one can easily obtain the canonically transformed Hamiltonian using Legendre tranformation. Thus, the final hamiltonian takes the form:

\begin{equation}
\mathcal{H}_0= N\left[\frac{P_x^2}{2}-\frac{P_y^2}{2}+(y^2-x^2)V(\phi)\right]\label{HClassicalFree}.
\end{equation}
The canonical transformation turns the Hamiltonian into 
 "ghost oscillator" for usual scalar fields which is expressed as the difference between two harmonic oscillators Hamiltonian.
The Eq.(\ref{HClassicalFree}) is purely classical and only expressed in different guises. Various methodologies exist for incorporating GUP. One approach involves deforming coordinates while momentum remains unchanged and the other is to deform momentum. In this paper, we adopt the latter approach where  we write the Hamiltonian after introducing momentum deformation using generalized gncertainty. To be specific, we introduce semi-classical canonical variables $q_i$ and $P_i$, and subsequently apply GUP on the Wheeler-DeWitt (WDW) equation of our cosmological model as follows: \cite{lopez2018phase,lopez2023generalized}.
\begin{equation}
q_i=q_{0i} ,\quad P_i=P_{0i}\left(1-\beta\gamma P_0 +2\gamma^2\frac{\beta^2+2\epsilon}{3}P_0^2\right),\label{momentumdeformation}
\end{equation}
where $P_0^2=P_{0j}P_{0j} $,  $\epsilon,\beta$, and $\gamma$ are the quantum-gravitational effect parameters.
Substituting Eq.(\ref{momentumdeformation}) in the original Hamiltonian Eq.(\ref{HClassicalFree})and taking terms only upto the order of 
$\gamma^2 $  gives:
\begin{widetext}
\begin{equation}
\mathcal{H}= \mathcal{H}_0  +\beta\gamma(P_{0y}^4-p_{0x}^4) +\gamma^2(P_{0x}^4-p_{0y}^4)\left(\frac{\beta^2}{6}+\frac{2\epsilon}{3}\right)+ \mathcal{O}(\gamma^3)\label{HgupFree},
\end{equation}
\end{widetext}
where $\mathcal{H}_0=\frac{P_{0x}^2}{2}-\frac{P_{0y}^2}{2}+(y^2-x^2)V(\phi)$ is the original Hamiltonian without GUP-distortion. Here subscript $0$ denotes the unperturbed Hamiltonian equation. And the unperturbed $x,y$ are denoted as $(q_{0x},q_{0y})$  while the corresponding unperturbed momentum pairs $(P_x, P_y)$ are indicated as $(P_{0x}, P_{0y})$.

Since we require  to re-express Eq.(\ref{HgupFree})in terms of cosmological variables. This can be achieved through the inverse transformation of variables in terms of scale factor, scalar field and there conjugate momenta. Using Eq.(\ref{xdot}) and Eq.(\ref{ydot}), we get
\begin{equation}
P_{0x} = \dot{x}= \frac{3}{2}\frac{a^{1/2}\dot a}{\mu} \sinh (\mu\phi)+(a^{3/2}\dot\phi) \cosh(\mu\phi),\label{Px}\\
\end{equation}
\begin{equation}
P_{0y}= \dot{y}= \frac{3}{2}\frac{a^{1/2}\dot a}{\mu} \cosh (\mu\phi) + (a^{3/2}\dot\phi) \sinh(\mu\phi).\label{Py}
\end{equation}\\

We choose $N(t)=1$ without loss of generality, as there is no dynamics in the lapse function.
$$\frac{\partial \mathcal{L}}{\partial \dot{a}}= P_a= -6\dot{a}a,$$ and $$ \frac{\partial \mathcal{L}}{\partial \dot{\phi}}= P_\phi= a^3\dot{\phi},$$
by substituting $P_a$ and $P_\phi$ in Eq.(\ref{Px}) the equations becomes
\begin{equation}
P_{0x}=-\frac{P_a}{4a^{1/2}\mu}\sinh(\mu\phi)+\frac{P_\phi}{a^{3/2}}\cosh(\mu\phi),\end{equation}
\begin{equation}
P_{0y}=-\frac{P_a}{4a^{1/2}\mu}\cosh(\mu\phi)+\frac{P_\phi}{a^{3/2}}\sinh(\mu\phi).
\end{equation}\\
Substituting  $P_{0x}$ and $P_{0y}$ in the momentum-deformed Hamiltonian Eq.(\ref{HgupFree}) by taking $\epsilon=1$ and $\beta=0$ gives,
\begin{widetext}

\begin{equation}
\mathcal{H}_{GUP}=-\frac{P_a^2}{12a}+\frac{P_\phi^2}{2a^3}+ V(\phi)a^3+ \alpha\left[\left((\frac{P_a^2}{12a}+\frac{P_{\phi}^2}{2a^3})\cosh(2\mu \phi)- \frac{P_aP_{\phi}}{4\mu a^2}\sinh{(2\mu \phi)}\right)\left(-\frac{P_a^2}{12a}+\frac{P_{\phi}^2}{2a^3}\right)\right].
\end{equation}

This is the required GUP distorted Hamiltonian in the cosmological phase space dominated by a quintessence scalar field with arbitrary potential up to second-order perturbation where $\alpha=8\gamma^2/3$. \\

Using Eq.(\ref{poisson})we can obtain the final Raychaudhuri equation:
\begin{eqnarray}
2\frac{\ddot{a}}{a}+\left(\frac{\dot a}{a}\right)^2 &=& -\frac{\dot\phi^2}{2}+V(\phi)+\alpha a^3 \Bigg[\Bigg((-H^2-\frac{\phi^2}{2})\cosh(2\mu \phi) + \frac{\sqrt{8}H \dot \phi}{\sqrt{3}}\sinh{(2\mu \phi)}\Bigg)\Bigg(-3H^2+\frac{\phi^2}{2}\Bigg) \nonumber\\
&& +\Bigg((3H^2+\frac{\phi^2}{2})\cosh(2\mu \phi) +\sqrt{6}H\dot \phi \sinh{(2\mu \phi)}\Bigg)\Bigg(H^2-\frac{\phi^2}{2}\Bigg)\Bigg]
\label{rayforlambda}.
\end{eqnarray}

Varying action with respect to $N$ gives the GUP distorted Hamiltonian constraint $\mathcal{H}'_{GUP}=0$, given in \cite{remmen2013attractor} gives
the final Friedmann equation as: 
\begin{equation}
3H^2=\frac{\dot \phi^2}{2}+ V(\phi)+ \alpha a^3\left[\left((3H^2+\frac{\dot \phi^2}{2})\cosh(2\mu \phi)+ \sqrt{6}H\sinh{(2\mu \phi)}\right)\left(-3H^2+\frac{\dot \phi^2}{2}\right)\right]\label{freidmann},
\end{equation}
and the K-G equation is
\begin{equation}
\Ddot{\phi}+3\dot\phi H + V,_{\phi}(\phi) = 
\sqrt{\frac{3}{2}} \alpha a^3\left[\left(3H^2+\frac{\phi^2}{2})\sinh{(2\mu \phi)} +\sqrt{6}H \dot \phi\cosh{(2\mu \phi)}\right)\left(3H^2-\frac{\dot \phi^2}{2}\right)\right] \label{klein}.
\end{equation}
\end{widetext}

With the effect of GUP-perturbation our original cosmological equations takes the form as Eq.(\ref{rayforlambda}), Eq.(\ref{freidmann})and Eq.(\ref{klein}). In all these equations the effect of GUP appears explicitly and by turning off the GUP term $\alpha$ we recovered the original equations.

Compared to the GUP-distortion in the phantom field in \cite{bh1} we observe that the phantom K--G equation remains the same after introducing GUP-distortion and shows no explicit GUP terms while it gets modified for quintessence.  
\section{Dynamical system analysis}\label{dsa}
In this section, we carry out stability analysis for our GUP-distorted system. Since our system is highly nonlinear and it is difficult to predict its behavior, the method of dynamical system analysis is used to extract qualitative behavior. In this method, we use a set of dimensionless variables that describe the state of the system in a finite phase space, and to understand the flow of these variables, we construct first-order autonomous differential equations. One can find fixed points by solving these coupled differential equations simultaneously, where a fixed point represents the points where overall dynamics vanish.
The nature of these fixed points is evaluated through the sign of eigenvalues from the Jacobian matrix at those points \cite{alho2022cosmological,s17,shahalam2015dynamics,Jaskirat,sh2,sh4}. 

We begin with the Friedmann equation as our starting point because it acts as a constraint equation. Next, we normalize it by the square of the Hubble parameter such that it brings all components contributing to the Hubble rate on an equal footing. This normalization ensures that each term becomes dimensionless, allowing us to express them in terms of dimensionless variables called Einstein-Nordstrom variables.

In the next section, we construct the autonomous differential equation for a given potential and perform a dynamical system analysis for the GUP-modified background cosmology. 

\subsection{ Potential $V(\phi)= V_0 \sinh^{-n}({\mu \phi})$}
In this subsection, we study qualitative dynamics
(DSA) for a universe dominated by a quintessence scalar field
with $V(\phi)= V_0 \sinh^{-n}({\mu \phi})$, considering GUP correction.
In order to determine the appropriate potential that fits with the evolution of the universe such that it shows asymptotic behavior like inverse power law in early time and exponential in the later phase
is the potential $V(\phi)= V_0 \sinh^{-n}({\mu \phi})$ as discussed in \cite{Phy1}. 

The normalized  Friedmann Eq.(\ref{freidmann}) takes the form as                      
\begin{widetext}
\begin{equation}
1= \frac{\dot \phi^2}{6H^2} +\frac{V(\phi)}{3H^2}+\frac{\alpha a^3 \dot \phi^4 \cosh(2\mu \phi)}{3H^2}- 3\alpha a^3 H^2 \cosh{(2 \mu \phi)} -\sqrt{6} \alpha a^3 H\dot \phi \sinh{(2\mu \phi)} +\frac{\alpha a^3 \dot \phi^3 \sinh{(2\mu \phi)}}{\sqrt{6}H} \label{dimon}.
\end{equation}
now defining new EN variables as
\begin{eqnarray}
x &\equiv& \sqrt{\frac{\dot \phi ^2}{6 H^2}}, \quad 
y \equiv \sqrt{\frac{V(\phi)}{3H^2}}, \quad 
z \equiv \sqrt{3 \alpha a^3 H^2 \cosh{(2\mu \phi)}}, \quad 
v \equiv \sqrt{\sqrt{6} \alpha a^3 \dot \phi H \sinh{(2\mu \phi)}}. 
\end{eqnarray}
and writing the Eq.(\ref{dimon}) in terms of  new variables is written as  
\begin{equation}
1= x^2 +y^2 + x^4 z^2 -z^2 - v^2 +x^2 v^2.
\end{equation}
Raychaudhuri's equation in terms of new variables is  
\begin{equation}
\frac{\dot H}{H^2}= \frac{3}{2}(-x^2+y^2-1)+z^2\left(\frac{2}{3}+x^2(2-3x^2)\right)-\left\{(1-\mu)x^2-\frac{1}{2}\right\}\left\{\frac{1-x^2-y^2-z^2(x^4-1)}{x^2-1}\right\} .
\end{equation}\\
\end{widetext}
As we have defined Raychaudhuri and Friedmann equation in terms of new variables we are ready to do the dynamical system analysis. To know the dynamics of the system we need the vector flow of each term as the dynamics evolve with time so firstly, we need to construct first-order differential equations for each new terms.  We are going to take slope of each terms with respect to newly defined time $N=\ln{a}$ where 'a' is the scale factor. However, constructing  autonomous differential equation alone with the help of new variables sometimes fails to close the system \cite{gong2014general,b18,ng2001applications,r14}. To close the system we need to define new variable $\lambda\equiv -\frac{V_{,\phi}}{V}$, 
\begin{eqnarray}
\lambda = \frac{n\mu}{\sinh{(\mu \phi)}}.
\end{eqnarray}
With this we construct the  autonomous equations:
\begin{widetext}
 \begin{eqnarray}
f(x,y,z)\equiv \frac{dx}{dN}&=&\left\{-\frac{3}{4}x^3+\frac{3}{4x}\right\}\left\{\frac{1-x^2-y^2-z^2(x^4-1)}{x^2-1}\right\}+3xz^2 +2\sqrt{6}x^3z^2-3x+\sqrt{\frac{3}{2}}\lambda y^2 
-x\Bigg[\frac{3}{2}(-x^2+y^2-1)\nonumber\\ &&+z^2\left(\frac{2}{3}+x^2(2-3x^2)\right)-\left\{(1-\mu)x^2-\frac{1}{2}\right\}\left\{\frac{1-x^2-y^2-z^2(x^4-1)}{x^2-1}\right\}\Bigg],\\
g(x,y,z)\equiv\frac{dy}{dN} &=& -\sqrt{\frac{3}{2}} \lambda x y
-y\Bigg[\frac{3}{2}(-x^2+y^2-1)\nonumber\\ &&+z^2\left(\frac{2}{3}+x^2(2-3x^2)\right)-\left\{(1-\mu)x^2-\frac{1}{2}\right\}\left\{\frac{1-x^2-y^2-z^2(x^4-1)}{x^2-1}\right\}\Bigg],\\
h(x,y,z)\equiv\frac{dz}{dN}&=& \frac{3}{2}z +\frac{3}{4z}\left\{\frac{1-x^2-y^2-z^2(x^4-1)}{x^2-1}\right\}
 +z\Bigg[\frac{3}{2}(-x^2+y^2-1)\nonumber\\ &&
 +z^2\left(\frac{2}{3}+x^2(2-3x^2)\right)-\left\{(1-\mu)x^2-\frac{1}{2}\right\}\left\{\frac{1-x^2-y^2-z^2(x^4-1)}{x^2-1}\right\}\Bigg], \\
u(x,y,z)\equiv\frac{d\lambda}{dN}&=& -\frac{(6\lambda x^2z^2)(x^2-1)}{{1-x^2-y^2-z^2(x^4-1)}}.
 \end{eqnarray}
\end{widetext}
\subsection{Fixed point analysis}
In this subsection, we perform dynamical system analysis for our chosen potential with the help of fixed points and linear stability theory to know the nature of each fixed point. Subsequently, we compare our GUP-distorted dynamics with the undistorted original dynamics.

To do this we first need to take the slopes of the above equations equal to zero to find the fixed point. These fixed points represent those points where our dynamics get static.  To know the behavior of fixed points we use the linear stability analysis method.

The analysis is shown in Table \ref{tablewith}.
As depicted from Table \ref{tablewith}  our dynamics get more richer than the original dynamics shown in Table \ref{tablewithout}. Incorporating the Generalized Uncertainty Principle(GUP)  introduce the addition of new variables into the original dynamics.
Since our dynamics consist of four variables, it necessitates the construction of a $4*4$ Jacobian matrix yielding four eigenvalues. The sign of these eigenvalues determines the nature of stability. 

In addition to that the fixed points also get shifted from the original dynamics and stable fixed points get vanishes. It means that with the introduction of GUP, our dynamics get modified with a comparable amount which is totally opposite behavior when we discussed GUP in the phantom field in \cite{bh1} where after the introduction of GUP dynamics get less richer and fixed points remain the same.\\

\begin{widetext}
\begin{table}[htbp]
\centering
\caption{\textbf{GUP-distorted Fixed points and the stability analysis for the potential $V(\phi)= V_0 \sinh^{-n}{\mu \phi}$}}
\resizebox{0.8\linewidth}{!}{
\begin{tabular}{l @{\hspace{5mm}} c @{\hspace{5mm}} c @{\hspace{5mm}} c @{\hspace{5mm}} c @{\hspace{5mm}} c @{\hspace{5mm}} c @{\hspace{5mm}} c @{\hspace{5mm}} c @{\hspace{5mm}} c}
\toprule
& \textbf{x} & \textbf{y} & \textbf{z} & \textbf{$\lambda$} & \textbf{$E_1$} & \textbf{$E_2$} & \textbf{$E_3$} & \textbf{$E_4$} & \textbf{Stability}\\
\midrule
\textbf{A} & $-0.2$ & $-1.01$ & $-4.8$ & $0$ & $340.6$ & $24.1$ & $-6.8$ & $0.2$ & saddle point \\
\textbf{B} & $-0.2$ & $1.01$ & $-4.8$ & $0$ & $340.6$ & $24.1$ & $-6.8$ & $0.2$ & saddle point \\
\textbf{C} & $-0.2$ & $-1.01$ & $4.8$ & $0$ & $340.6$ & $24.1$ & $-6.8$ & $0.2$ & saddle point \\
\textbf{D} & $-0.2$ & $1.01$ & $4.8$ & $0$ & $340.6$ & $24.1$ & $-6.8$ & $0.2$ & saddle point \\
\textbf{E} & $0.2$ & $-1.01$ & $-4.8$ & $0$ & $340.6$ & $24.1$ & $-6.8$ & $0.2$ & saddle point \\
\textbf{F} & $0.2$ & $1.01$ & $-4.8$ & $0$ & $340.6$ & $24.1$ & $-6.8$ & $0.2$ & saddle point \\
\textbf{G} & $0.2$ & $-1.01$ & $4.8$ & $0$ & $340.6$ & $24.1$ & $-6.8$ & $0.2$ & saddle point \\
\textbf{H} & $0.2$ & $1.01$ & $4.8$ & $0$ & $340.6$ & $24.1$ & $-6.8$ & $0.2$ & saddle point \\
\bottomrule
\end{tabular}
}
\label{tablewith}
\end{table}

\begin{table}[htbp]
\centering
\caption{\textbf{without GUP-distorted Fixed points and the stability analysis for the potential $V(\phi)= V_0 \sinh^{-n}{\mu \phi}$}}
\resizebox{0.6\linewidth}{!}{
\begin{tabular}{l @{\hspace{5mm}} c @{\hspace{5mm}} c @{\hspace{5mm}} c @{\hspace{5mm}} c @{\hspace{5mm}} c @{\hspace{5mm}} c  @{\hspace{5mm}} c}
\toprule
& \textbf{x} & \textbf{y} & \textbf{$\lambda$} & \textbf{$E_1$} & \textbf{$E_2$} & \textbf{$E_3$} &  \textbf{Stability}\\
\midrule
\textbf{A} & $0$ & $0$ & $c$ & $-3/2$ & $3/2$ & $0$  & saddle point \\
\textbf{B} & $-1$ & $0$ & $0$ & $3$ & $3$ & $3/2$  & unstable point \\
\textbf{C} &$1$ & $0$ & $0$ & $3$ & $3$ & $-3/2$  & saddle point \\
\textbf{D} & $0$ & $-1$ & $0$ & $-3$ & $-3$ & $0$  & stable point \\
\textbf{E} & $0$ & $1$ & $0$ & $-3$ & $-3$ & $0$  & stable point \\
\bottomrule
\end{tabular}
}
\label{tablewithout}
\end{table}
\end{widetext}

On comparing Table \ref{tablewithout} and Table \ref{tablewith} we observe that the points A, B, and C vanishes in the GUP-modified dynamics. However, points D and E get distorted from the original points (0,1) and (0,-1) with some factors. These points in the GUP-distorted Table \ref{tablewith} show the spectrum of points because of the addition of one more variable into the dynamics. In physical terms, within the original dynamics, points D= (0,-1) and E=(0,1) show stability means the dynamics settle down for zero velocity with positive and negative field values. However, the introduction of GUP perturbs the dynamics causing distortion in the values of fixed point such that it alters the stability.

\section{Conclusion}
In this paper, we construct the GUP-modified Hamiltonian starting from the Einstein--Hilbert action for quintessence scalar field. In 
the first section \ref{classicaltreatment},  we review the cosmological equations obtained from the classical Hamiltonian.  In the subsequent subsection \ref{GUP deformed},  we introduce phase space variables $x$ and $y$ that encode information about our cosmological variable.  We express our cosmological Hamiltonian in terms of these variables to maintain simplicity. Then, we directly incorporate momentum-deformed GUP into the newly obtained Hamiltonian using the WDW equation for our cosmological model. Consequently, we derived the cosmological equations Friedmann, Raychaudhuri, and Klein-Gordon equation from the GUP-distorted Hamiltonian.
As a result, we see GUP-modified term explicitly in Eq.(\ref{rayforlambda}), Eq.(\ref{klein}), and Eq.(\ref{freidmann}) which is entirely different from the phantom fields, where the K--G equation does not exhibit explicit dependence. 

To observe the impact of GUP-distortion on the dynamics of our universe, we conduct a dynamical system analysis using the potential $V(\phi) = V_0 \sinh^{-n}{\mu \phi}$  in section (\ref{dsa}). Additionally, we compare the GUP-distorted dynamics with the original dynamics in the table (\ref{tablewith}) and (\ref{tablewithout}).
From the table (\ref{tablewith}) and (\ref{tablewithout}) we observe that our original dynamics also get modified with the introduction of GUP. Fixed points $A, B,$ and $C$ in table \ref{tablewithout} vanishes in table \ref{tablewith} with the effect of GUP-perturbation. Points $D$ and $E$ get distorted from their original position with some small amounts and stability also gets changed. With GUP modification the dynamics depend on one more variable that enriches the overall system. 

In the future, we intend to expand our analysis to encompass more generalized models, which incorporate anisotropic spacetime alongside anisotropic sources. Specifically, we find it compelling to explore along the lines presented in reference \cite{al1}, and to ascertain the constraints on the parameters therein for a realistic scenario within a GUP-modified setup. This endeavor is left as a project for our future pursuits.


\begin{acknowledgments}A.W. is partly supported by the US NSF  grant, PHY-2308845.
\end{acknowledgments}



\end{document}